\pdfoutput=1
\documentclass{ws-jai}
\usepackage[flushleft]{threeparttable}
\usepackage{graphicx}
\usepackage{lipsum}
\usepackage{float}
\usepackage{amsmath}
\usepackage{latexsym}
\usepackage{textcomp}
\usepackage{xcolor}
\usepackage{booktabs,etoolbox}
\usepackage{url}

\usepackage{silence}
\newlength{\toprulewidth}
\patchcmd{\toprule}% <cmd>
{\heavyrulewidth}{\toprulewidth}% <search><replace>
{}{}% <success><failure>
\begin{document}
\catchline{}{}{}{}{}

\markboth{S.~Ambily}{Development of Data Acquisition Methods for an FPGA-Based Photon Counting Detector}

\title{Development of Data Acquisition Methods for an FPGA-Based Photon Counting Detector}

\author{S. Ambily $^\dagger$, Mayuresh Sarpotdar, Joice Mathew, A. G. Sreejith, K. Nirmal, Ajin Prakash, Margarita Safonova and Jayant Murthy}

\address{Indian Institute of Astrophysics, Bangalore, India}

\maketitle

\corres{$^\dagger$Corresponding author. E-mail: ambily@iiap.res.in}

% \begin{history}
% \received{(to be inserted by publisher)};
% \revised{(to be inserted by publisher)};
% \accepted{(to be inserted by publisher)};
% \end{history}

\begin{abstract}
MCP-based detectors are widely used in the ultraviolet (UV) region due to their low noise levels, high sensitivity and good spatial and temporal resolution. We have developed a compact near-UV (NUV) detector for high-altitude balloon and space flights, using off-the-shelf MCP, CMOS sensor, and optics. The detector is designed to be capable of working in the direct frame transfer mode as well in the photon-counting mode for single photon event detection. The identification and centroiding of each photon event are done using an FPGA-based data acquisition and real-time processing system. In this paper, we discuss various algorithms and methods used in both operating modes, as well as their implementation on the hardware. 
\end{abstract}

%\keywords{Detectors, photon counting, FPGA, CMOS, centroiding, algorithms, balloon}

%%%%%%%%%%%%%%%%%%%%%%%%%%%%%%%%%%%%%%%%%%%%%%%%%%%%%%%%%%%%%
\section{Introduction}
\label{sec:intro} % \label{} allows reference to this section

Access to space has become easier and cheaper with the advent of small satellites such as e.g. Cubesats \cite[e.g][]{Noah_Cubesat}. We are developing compact scientific payloads using off-the-shelf components with tight constraints on weight, size and power, for atmospheric and astronomical studies in the NUV range (200--350 nm). We test and qualify these payloads using high-altitude balloons, with the expectation of space flights when the opportunity arises. In particular, we have designed and developed the MCP-based photon counting detector which we plan to fly in the near future. 

Photon-counting detectors using MCPs have been the standard for UV payloads because of their low readout noise, large detector area, radiation tolerance and long-wavelength rejection \cite{Kimble}. There are variations in the number of the MCP stacks, types of filters and photocathodes, and in the anode and readout electronics \cite{Joseph}. For complex payloads with high levels of customization, all-electronic readout mechanisms, such as Cross Strip (XS) anodes, Wedge and Strip anodes, Multi-Anode Microchannel Arrays (MAMA), or Delay-Line anodes are used \cite{Vallerga_MCP_detectors}. The charge clouds from the MCP electrodes in these system are read out by a custom electronics board which converts the amplified electron charge cloud into a corresponding voltage level \cite{Vallerga_2kx2k_MCP}. These readouts are often expensive and, while their cost is justified for high impact missions such as {\it GALEX} \citep{Galex_Detectors} or {\it FUSE} \citep{Fuse_Detector}, they may not be needed in less ambitious missions.

We have achieved significant savings in cost by using a phosphor-screen anode as the readout \cite[e.g.][]{Bonanno_ICMOS}, and by focusing the image of the anode onto a CMOS sensor using a lens. We have chosen a photocathode (S20) with sensitivity in the range 200--900 nm for our initial development, so that we can use it in both the NUV and the visible. The detector works as a CMOS camera in bright light, transferring each frame serially to the data processing system, and as a photon-counting detector in faint light conditions. In both photon counting and continuous frame-readout modes, the data acquisition and processing unit is the same field-programmable gate array (FPGA) board. Therefore, the modes can be changed by just switching from one code to another on the FPGA, without the necessity of making changes to the hardware. 

In this paper, we describe the various aspects of the design and implementation of the detector, with a focus on the data acquisition algorithms that run on the FPGA. We also discuss the hardware implementation of these algorithms, as well as the results of few calibration and characterization tests on the whole sensor. 

\section{Detector Overview}

\begin{figure}[!ht]
\begin{center}
\includegraphics[width=180mm]{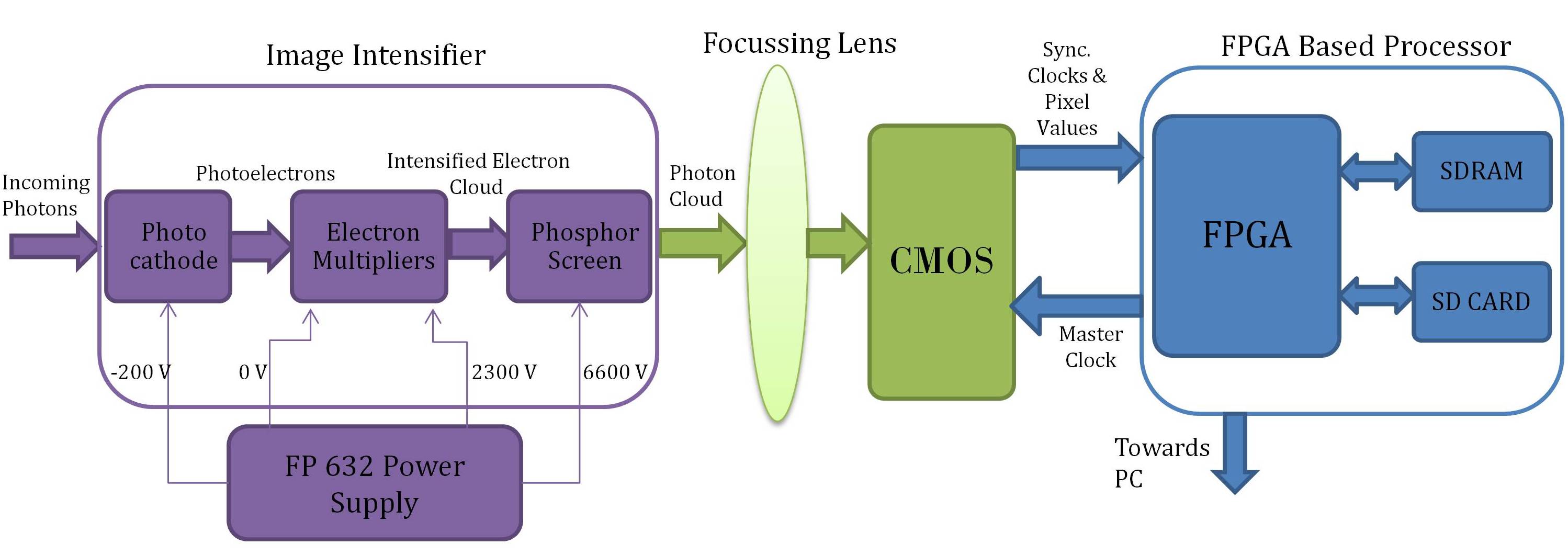} %100 percent
\end{center}
\caption{Detector Block Diagram}
\label{fig:blockdia}
\end{figure}

The detector (Fig.~\ref{fig:blockdia}) is an intensified CMOS camera comprising the two parts:
\begin{enumerate}
\item An image intensifier with an MCP and its associated high-voltage power supply (HVPS);
\item A readout and processing unit, which includes a relay lens, CMOS sensor and a digital readout card using an FPGA.
\end{enumerate} 
  
% Detector Blocks
\subsection{Image Intensifier}

We are using an MCP340 assembly from  Photek\footnote{\tt{http://www.photek.com/}}. It is a 40 mm, Z-stack MCP with an S20 photocathode deposited on a quartz input window and P46 phosphor deposited on the anode. The distance from the photocathode to the MCP front surface is 100 microns and the distance from the MCP output surface to the phosphor is 700 microns.

The MCP is powered by the FP632 micro-HVPS from Photek which weighs only 100 gm, ideal for our lightweight ($< 3$ kg) balloons. The HVPS provides several outputs with voltages from $-200$ V to $6,900$ V. The gain of the MCP is set by the voltage at the MCP output, which automatically adjusts the screen voltage, maintaining a constant MCP-out-Anode voltage difference. The resulting electron cloud from the MCP output electrode is accelerated towards the phosphor screen anode (with a peak response at 530 nm) and the resulting photons are focused on the CMOS surface by a relay lens. 

\subsection{Relay Optics}

Electrons from the MCP induce fluorescence in the P46 phosphor screen (40-mm diameter) which is imaged onto the CMOS readout (3.4-mm diameter) using relay optics. Similar commercial systems often use a fiber-optic coupling but these are more difficult to physically implement with high precision. We are now using the  TVF-3.6X2812IR-BCDN-MD, a board-level varifocal lens from Senko with high precision zoom and focus adjustment which was provided with the CMOS board. The lens system has an $F$-ratio of $F1.2$ with an adjustable focus from 3 mm to 10 mm. 

This system introduces a significant amount of aberrations into the system because of the large demagnification factor from the MCP to CMOS (17:1) and we are designing a new lens system to better couple the two. 

\subsection{CMOS Sensor}
We have chosen a 1 mega-pixel video image sensor, OV9215 from OmniVision Technologies\footnote{\tt{http://www.ovt.com/}}, for the CMOS camera. It is a 1/4 inch sensor that provides full-frame, sub-sampled or windowed 8-bit/10-bit images. The advantages of CMOS sensors in easier readout and circuit integration, as well as their low power requirements, make them an obvious choice over CCDs for our applications. The disadvantage of CMOS sensors is that they have an uncertain photometric flat field but this is not relevant in our photon counting context\cite{Uslenghi_ICMOS}.  The CMOS chip sits on an off-the-shelf headboard, originally designed for use in CCTV applications\footnote{Soliton Technologies: http://http://www.solitontech.com/}, which provides a bias and clock for the sensor. The headboard reads the digital image values and sends them to the FPGA for further processing. It also incorporates a synchronous serial port, which may be used to modify the internal camera registers when needed. 
The pixel stream and the three sync signals for the pixel, line, and frame clocks are generated by the CMOS sensor and are connected directly to the FPGA inputs. The master input clock for the CMOS chip is generated by the on-chip crystal oscillator and PLL on the FPGA board. The oscillator frequency is fixed at 12 MHz, but the input clock frequency for the CMOS chip may be varied from 6 to 27 MHz using the programmable PLLs on the FPGA. The successive pixel values are read out as a 2D image matrix and stored in an array on the FPGA memory.  The limiting factor in the frame rate is the CMOS frequency, and we are in the process of obtaining and testing faster chips.
 
\subsection{Data Acquisition Board}

We have chosen the XuLA2-LX25 FPGA prototyping board from XESS Corp.\footnote{\tt{http://www.xess.com/shop/product/xula2-lx25/}} for the design of the main processor since the often used commercial development boards require too many resources (power, weight, complexity). This board meets our minimum requirements in a compact form factor. The heart of the FPGA board is a Spartan\textsuperscript{\textregistered} 6 series XC6LX25 chip\footnote{Xilinx, Inc: \tt{http://www.xilinx.com/}}, which controls and reads out the CMOS sensor, processes the data, stores and transmits the output. The use of FPGAs allows us to rapidly prototype different algorithms without changing the hardware which is invaluable in a development environment.

The FPGA modules were written using Verilog and synthesized and programmed using the Xilinx ISE\textsuperscript{\textregistered} (Integrated Synthesis Environment) Design Suite. In addition, we have used standard VHDL modules\footnote{XESS Corp.} for interfacing the SDRAM (synchronous dynamic random-access memory), PLL (Phase Locked Loop), and micro-SD (Secure Digital) card to the Spartan\textsuperscript{\textregistered} 6 chip on the FPGA board. 

\section{Implementation}

The detector is designed to operate in two modes: 
\begin{itemize}
\item Continuous frame transfer mode.
\item Photon counting mode.
\end{itemize}

We have designed the readout and associated electronics of the detector so that we can switch between both modes with no changes in the hardware. Incoming photons eject electrons from the S20 photocathode which are multiplied by the MCP into a charge pulse that falls on a green phosphor screen. The emitted photons are read by the CMOS sensor with further processing done by the FPGA \citep{{Bellis_ICMOS},{Fidouh}}.

The overall implementation of the detector has been broken down to (a) development of the front end photonics system, and (b) interfacing of the digital backend electronics. The first task involves coupling the light from the MCP to the CMOS sensor through a relay lens system with minimum distortion and light loss. The second step involves the clocking and synchronization of the CMOS chip, as well as the programming and interfacing of the FPGA with the host PC. This part is further divided into four steps, which are described in detail in the following subsections.

\subsection{Centroid calculation}

Any photon event on the photocathode creates a charge cloud that is Gaussian in shape. This cloud is spread over multiple pixels on the sensor which we centroid with sub-pixel accuracy using the FPGA. The centroiding algorithm was implemented using a $3\times 3$ pixel sliding window during our tests which we later expanded to a $5\times 5$ pixel window \cite{Hutchings}. This yielded better accuracy within our computational limits. 

\begin{figure}[ht!]
\begin{center}
\includegraphics[scale=0.12]{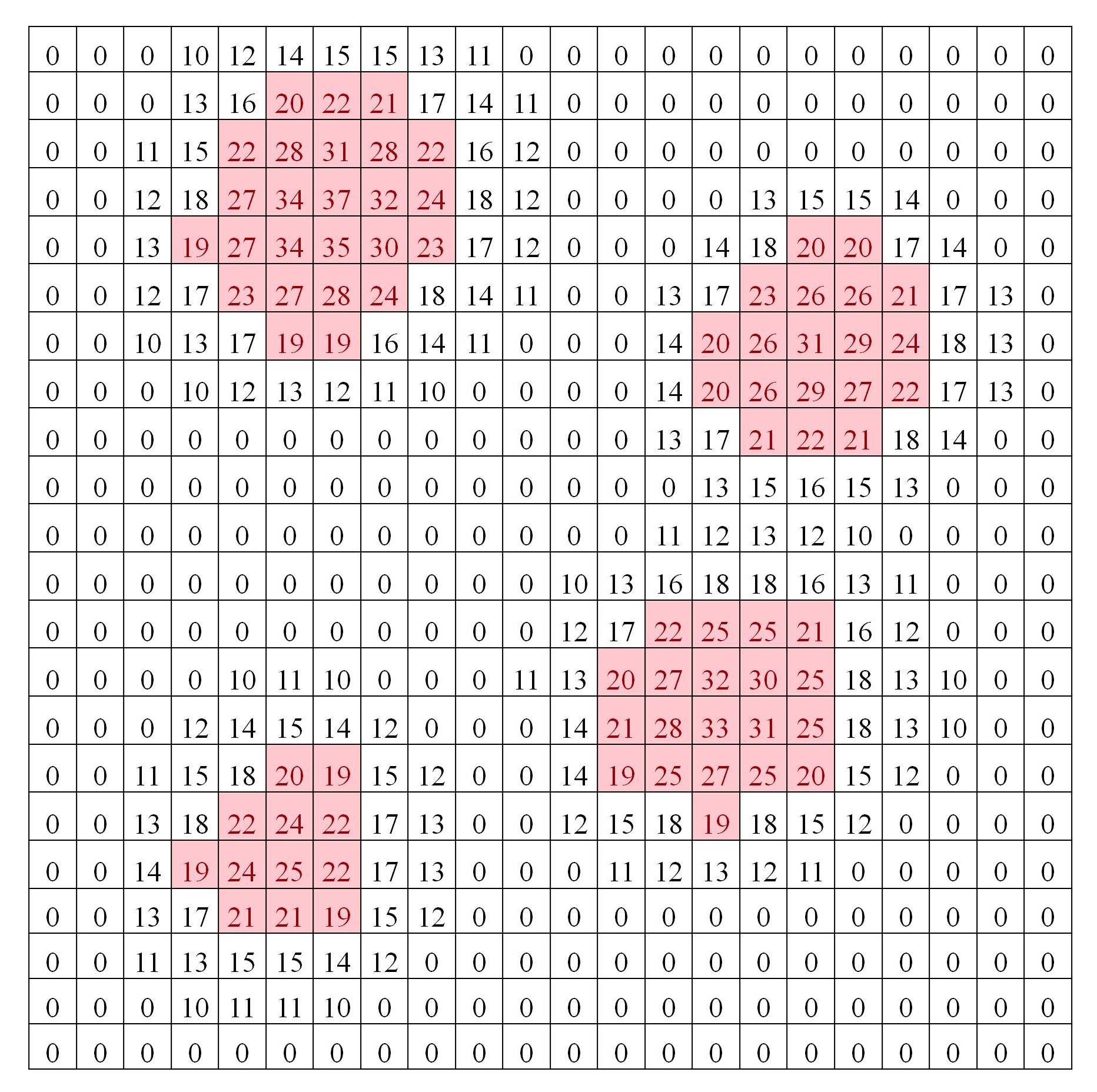}
\includegraphics[scale=0.12]{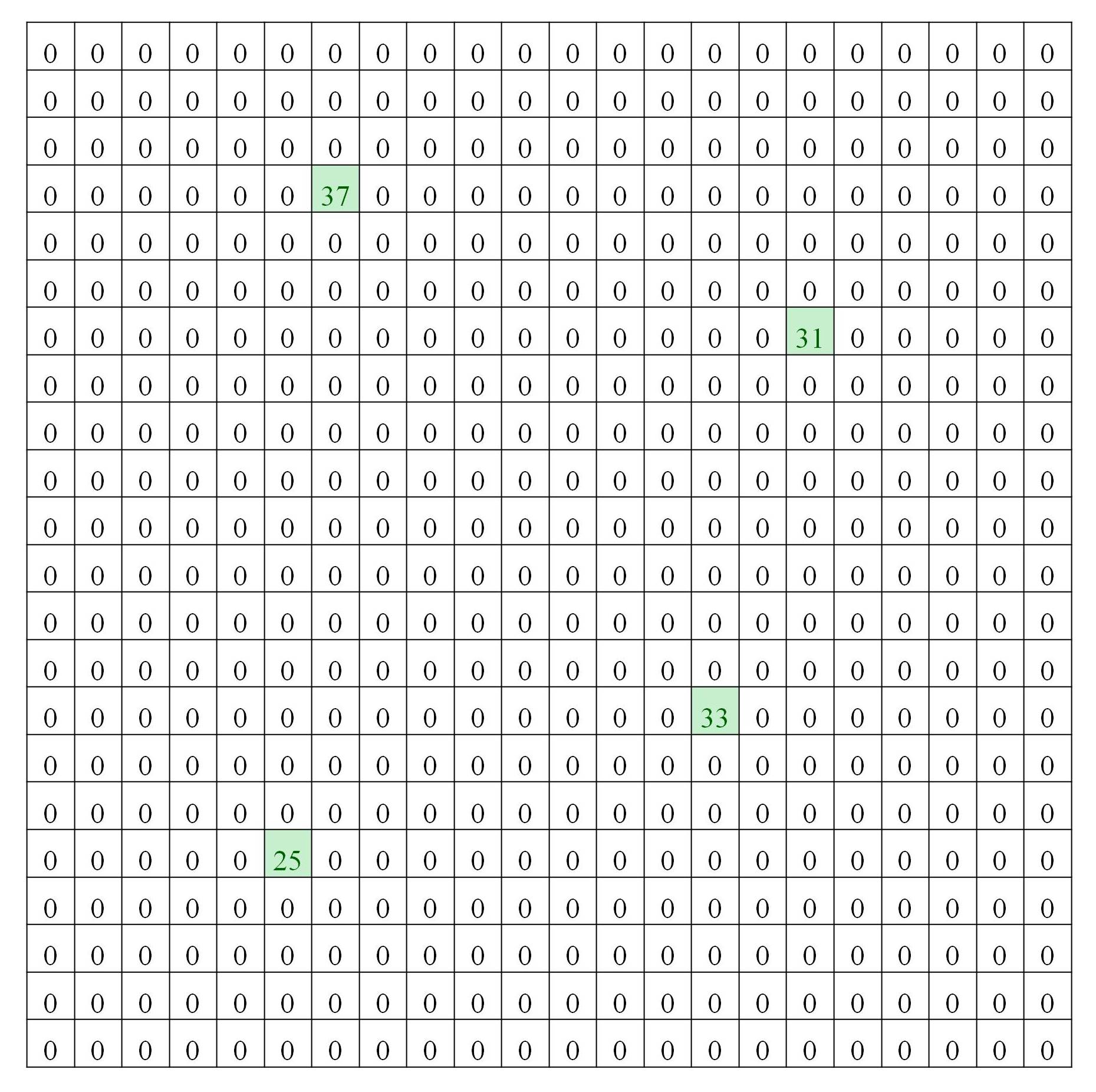}
\end{center}
\caption{{\it Left}: Simulated image array. The values retained after a threshold level are highlighted and the rest of the values are considered to be zero. {\it Right}: Reconstructed image.}
\label{fig:matrix1}
\end{figure}

If we operate in the continuous frame transfer mode, all the pixel values will be stored on the SDRAM module which will rapidly exceed our on-board storage. Centroiding on the chip is space-efficient but can be computationally expensive. We implemented a scheme in which we stored only the most recent five rows of the current frame on the FPGA chip ($5 \times 1280$ pixels). This is illustrated in (Fig.~\ref{fig:matrix1}, {\it Left}) where we have simulated the sample image array in MATLAB\textsuperscript{\textregistered}\footnote{MathWorks, Inc.: \tt{http:in.mathworks.com/products/matlab/}}. The image consists of $25\times25$ pixels and the pixels were read out serially to a $5\times25$ array. The initial values of the array were assigned to be zero. As each pixel row from the sample image was read, the last row of the array on the simulated FPGA was updated, while the older rows were overwritten with the latest values. This sequence is illustrated in Fig.~\ref{fig:row1}. The flow chart for the centroid calculation steps is given in Appendix~A.\\

\begin{figure}[ht!]
\begin{center}
\includegraphics[width=150mm]{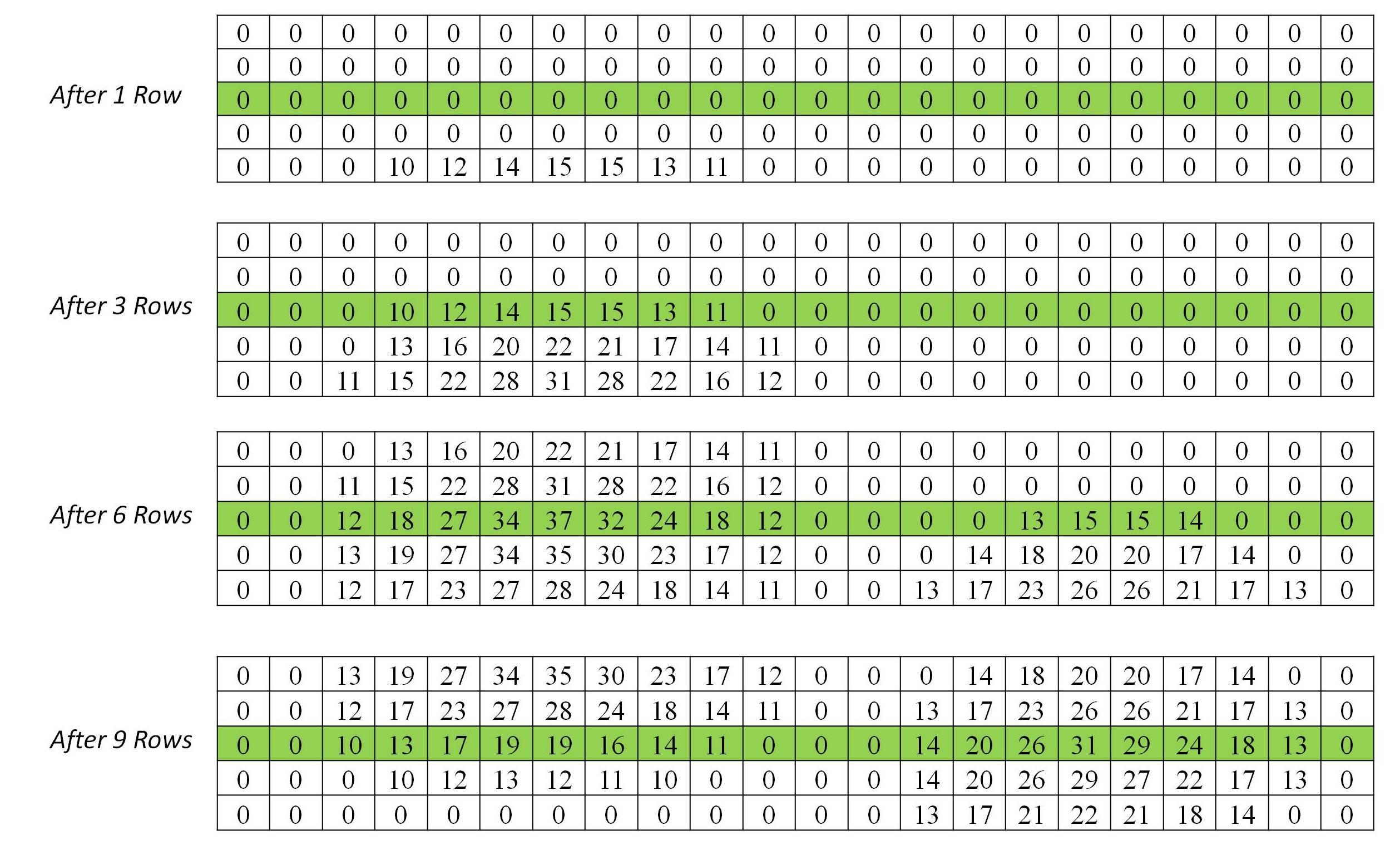} %100 percent
\end{center}
\caption{The implemented readout: the latest rows are shifted to the FPGA array in a pixel to pixel manner. }
\label{fig:row1}
\end{figure}

\subsubsection{Thresholding and detecting the local intensity maxima}

The first part of our centroiding procedure is to eliminate the dark noise arising from the S20 photocathode, which was measured in the lab (see Sec.~\ref{sec:dark}) to be 1000 cps at 23$^{\circ}$C. We dynamically estimate the local background level for each photon event, and use this value as the event threshold. This is done by taking the lowest corner pixel value in the centroiding window \cite{Hutchings}. 

As and when each pixel is read out, the array containing the last five rows is updated while the older row values are overwritten. The 24 pixels surrounding the last read-out pixel are read into a separate 5$\times$5 window array, and the window is checked for the presence of any local maxima. We flag the hot pixels and multiple events before calculating the centroids. Hot pixels are removed by eliminating any event where only the central pixel is higher than the threshold. Multiple events are marked by adding the maximum-minimum corner values to the output centroid packets \cite{Hutchings}.

\subsubsection{Calculating the centroid around the maximum points}

\begin{figure}[ht!]
\begin{center}
\includegraphics[width=180mm]{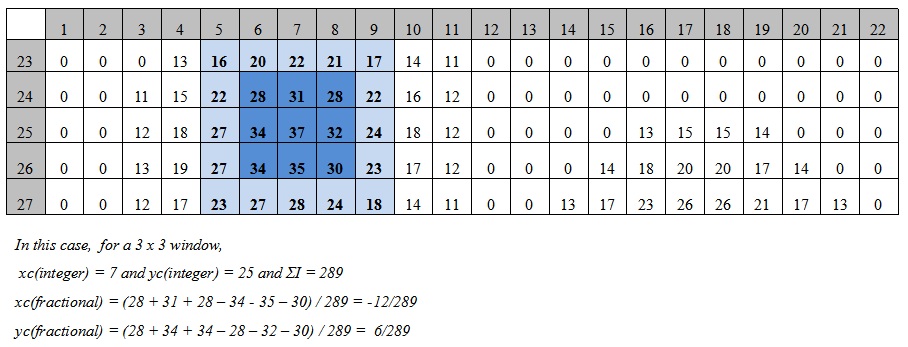} %100 percent
\end{center}
\caption{Centroid computation from each of the windows. The values obtained here are directly stored as a packet and fractional pixel levels of up to 8-bit accuracy are calculated at a later stage.}
\label{fig:row2}
\end{figure}

Once we have identified the hot pixels and multiple events, we calculate the centroids for the actual photon events in the current row. We saved the coordinates of the central pixel as the integer parts of the centroid. We computed the sub-pixel values in parallel by calculating the difference in intensities between the neighboring top and bottom pixels and the left and right pixels of the event (Fig.~\ref{fig:row2}). Once calculated, each centroid value consists of the $x$ and $y$ coordinates with their integer and fractional parts and a sign bit to indicate the direction of actual centroid coordinate (top/bottom and left/right) from the central pixel coordinate.

\subsection{Data Storage and Transmission}
\subsubsection{Photon Counting Mode}

\begin{figure}[ht!]
\begin{center}
\includegraphics[width=180mm]{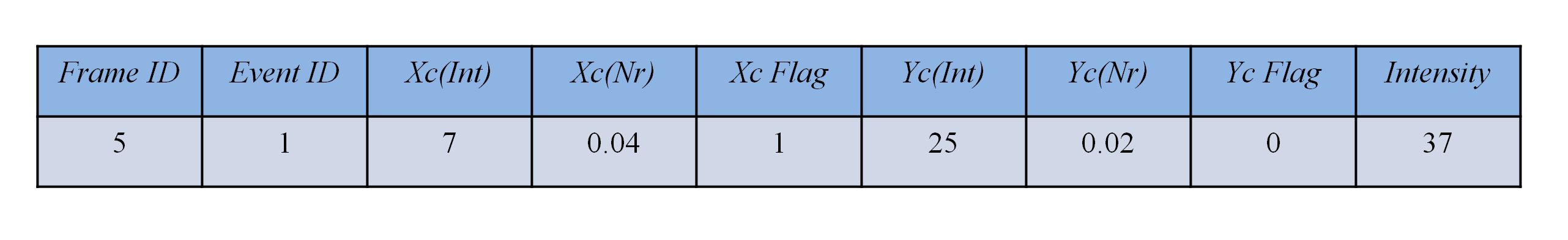} %100 percent
\end{center}
\caption{Data packet for each photon event. The final output packet is optimized to contain 7 or 8 bytes depending upon the sub-pixel accuracy required.}
\label{fig:row3}
\end{figure}

In the centroiding mode, the output data for telemetry is in the form of packets. Each packet contains a frame ID, event ID, integer and fractional values of centroid coordinates, and the intensity value of the central pixel (Fig.~\ref{fig:row3}). These packets are saved on the FPGA by creating a block of registers on the chip. This results in a faster operation: the packets from one frame can be sent over the serial port while the next frame is read. However, the number of photon events that can be saved per frame is limited by the number of logic cells available on-chip. In order to solve this, we save each packet to the SDRAM as and when it is generated.

Hardware description languages do not support division operation as it is impossible to realize a divider module in FPGA hardware and division operations are accomplished using repeated subtractions which add cost to the time and resources. Therefore, when transmitting the packet from the centroiding to the telemetry module, the sub-pixel values are not computed but are simply sent as the numerator and the denominator values. Before transmission, the sub-pixel values can be calculated with an accuracy of 4 bits each by a separate divider module. Because the RS232 transmission runs at a much slower speed than the FPGA clock, the telemetry unit runs the divisions in parallel and transmits the 4-bit fractional values for each packet. It is possible to improve the accuracy to 6 or 8 bits but at the expense of additional clock cycles and hardware resources. 

We have implemented an SD card interface for in-flight experiments, where instead of sending the data packets to the PC using serial port, we use a much faster Serial Peripheral Interface (SPI) protocol to store the data packets on the micro-SD card on the FPGA board. The values from the SD card can be read out as a text file or through the FPGA serial port communication at a later stage. 

\subsubsection{Continuous Frame Transfer Mode}

In this mode, the pixel values are directly transferred from the SDRAM to the SD card on-board after each frame is read. Since the SDRAM data width is 16 bits and our images are of 8 bits or 10 bits, the remaining bits are used to store the frame and row number which help in basic sanity checks of the data (such as, check for missing frames, incomplete rows, etc.). 

\subsection{Image Reconstruction}

Image reconstruction is intended to derive a meaningful image from the data packets and is done on the host PC (a single board computer in the case of a balloon flight) which receives the serial data from the FPGA board. In continuous imaging mode, the pixel values that are read from the SD card are stored into an image array after basic sanity checks. In the case of photon counting mode, the data stream itself contains a time stamp/frame ID, as well as the $x,y$ coordinates and intensity information, which makes the image reconstruction easier. For example, in Fig.~\ref{fig:matrix1} ({\it Right}) we display the image reconstructed from the centroids, calculated from the simulated image (Fig.~\ref{fig:matrix1}, {\it Left}, in Section 3.2). The image reconstruction flow chart is given in Appendix~B.

\section{Simulations}
We have tested the algorithm in MATLAB\textsuperscript{\textregistered} using a sample image array of $256\times 256$, with simulated Gaussian photon events of random dimensions, positions, and intensity. Poisson noise was added to the image, where the output pixel intensity was generated from a Poisson distribution with mean equal to the input pixel value. In addition, salt and pepper noise was added to this image with a density of 2\%. The image values were normalized to a range between 0 and 1 to simplify the programming. Pixel values from the image array were read serially in the same way as the CMOS chip readout. 

\begin{figure}[!ht]
\centering
% \begin{minipage}{.5\textwidth}
%   \centering
  \includegraphics[width=60mm]{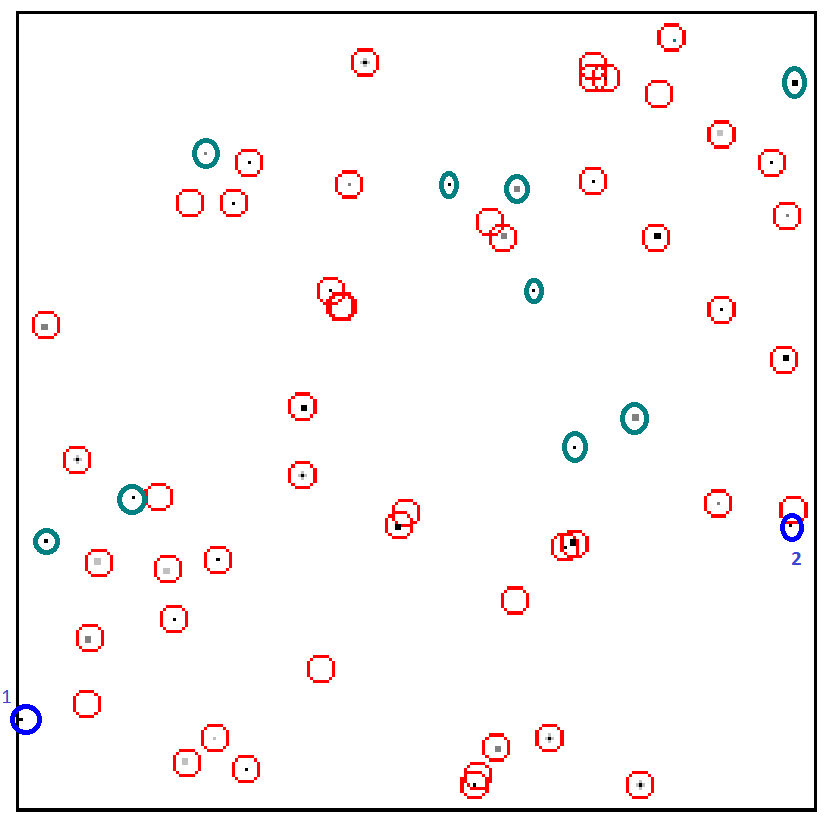}
 \hskip 0.3in
%  \end{minipage}%
% \begin{minipage}{.5\textwidth}
%   \centering
  \includegraphics[width=60mm]{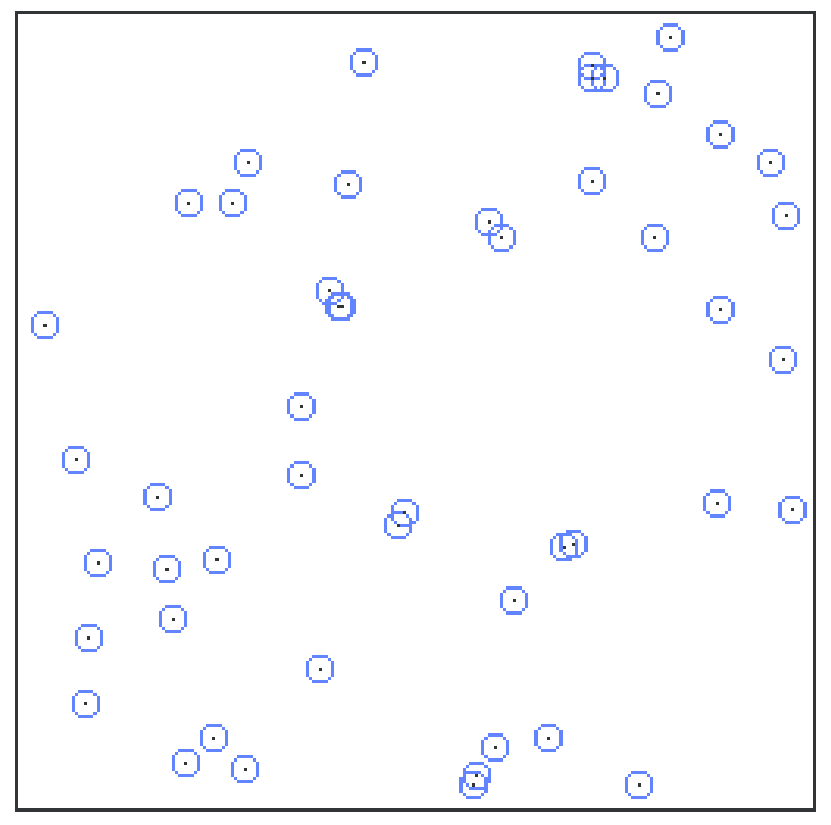}
% \end{minipage}
\caption{{\it Left}: Original image with the detected centroids highlighted. {\it Right}: Image after centroiding and reconstruction with a threshold value of 0.0025. The photon events are shown by circles, detected centroids are circled in red. Green circles represent the isolated hot pixels, and the blue circles are the events that were undetected (numbered 1 and 2).} 
\label{fig:result1}
\end{figure}

We have run a number of simulations and found that the mean number of photon events in the simulated array was 47.3 with an SNR of 14.6. The centroiding algorithm with a $5\times5$ window and different levels of thresholding was applied on the sample image to see how the threshold value affects the detection efficiency and noise rejection in the output image. We have illustrated this for one case which had 49 photon events in Fig.~\ref{fig:result1} ({\it Left}), with the actual photon events circled in red and the isolated hot pixels in green. The blue circles show the points that are actual events but were undetected in the reconstructed image due to their proximity to the edges of the image (numbered 1 and 2). The reconstructed image with the threshold value equal to the mean noise level of the image (0.0025 in this case) is shown in Fig.~\ref{fig:result1} ({\it Right}). In Table~\ref{table:compare_matlab} we show the differences in the number of detected events as the threshold values are changed. We see that the optimum threshold value has to be at, or slightly higher than, the mean noise level of the original image for detecting the maximum number of photon events with higher signal-to-noise ratio (SNR).

\begin{table}[!ht]
\begin{center}
%\parbox{.45\linewidth}{
\caption{Comparison of simulated images after reconstruction. First column displays the threshold value applied on the original image. Second column shows the number of detected photon events after reconstruction. The last column displays the SNR after reconstruction.} 
\vskip 0.1in
\begin{tabular}{|l|c|l|} 
\hline
\rule[-1ex]{0pt}{3.5ex} Threshold value & Number of detected events  & SNR \\
\hline
\rule[-1ex]{0pt}{3.5ex} 0.0 & 66 &  14.9868   \\
\rule[-1ex]{0pt}{3.5ex} 0.001 & 58 &     15.3433  \\
\rule[-1ex]{0pt}{3.5ex} 0.002 & 54  &     15.5892   \\
\rule[-1ex]{0pt}{3.5ex} 0.0025 (mean signal value) & 51  &     15.6330  \\
\rule[-1ex]{0pt}{3.5ex} 0.004  & 51 &     15.6330    \\
\hline
\end{tabular}
\label{table:compare_matlab}
\end{center}
\end{table} 

\begin{figure}[!ht]
\begin{center}
\includegraphics[width=60mm]{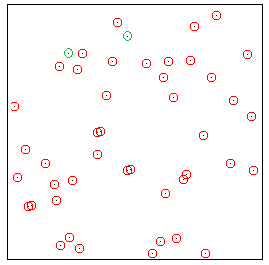} %100 percent
\end{center}
\caption{Results from the FPGA implementation. Image after centroiding and reconstruction. The photon events are shown by circles, detected events are circled in red, and green circles are the hot pixels.}
\label{fig:result2}
\end{figure}

The second step was to implement and test the algorithm in the FPGA hardware. The intensity values of the same sample image (Fig.~\ref{fig:result1}, {\it Left}) were approximated to 8-bit values ($0 -255$) due to the space constraints in the FPGA memory. The values were serially read by the centroiding module with the frame and line valid signals from the CMOS chip with the mean noise level as the threshold. The centroid coordinates were transferred to a PC through a serial-to-USB converter. The reconstructed image (in the same case as discussed above) has an SNR of 15.72 and 47 detected photon events, of which 45 are actual events and 2 are hot pixels. The fewer number of detected events (47 vs. 49) were because of the 8-bit approximation applied to the original image. The reconstructed image from the FPGA is shown in  Fig.~\ref{fig:result2}, with detected events highlighted in red and hot pixels in green. 

\section{Hardware Implementation and Characterization}
\label{sec:dark}

We are operating the MCP with the gain of $10^6$ for the typical voltage values of $-200$, $2200$, and $6000$ V at the cathode, MCPOut, and Anode terminals, respectively. The whole detector is powered by an in-house made power supply board which houses the FP632 power supply for the MCP, 5 V LiPo battery, and a voltage regulator. The distance between the phosphor screen and the front of the relay optics is currently kept at 3.5 cm, with the option of fine adjustments by using shims of about $100\,\mu$m. This distance was arrived at after a few trials, to provide the best focus for imaging the entire 40 mm diameter of the MCP on the CMOS surface. The CMOS is fixed at the focal plane of the relay lens at 10 mm distance. 

\begin{figure}[ht!]
%\centering
\includegraphics[height=60mm]{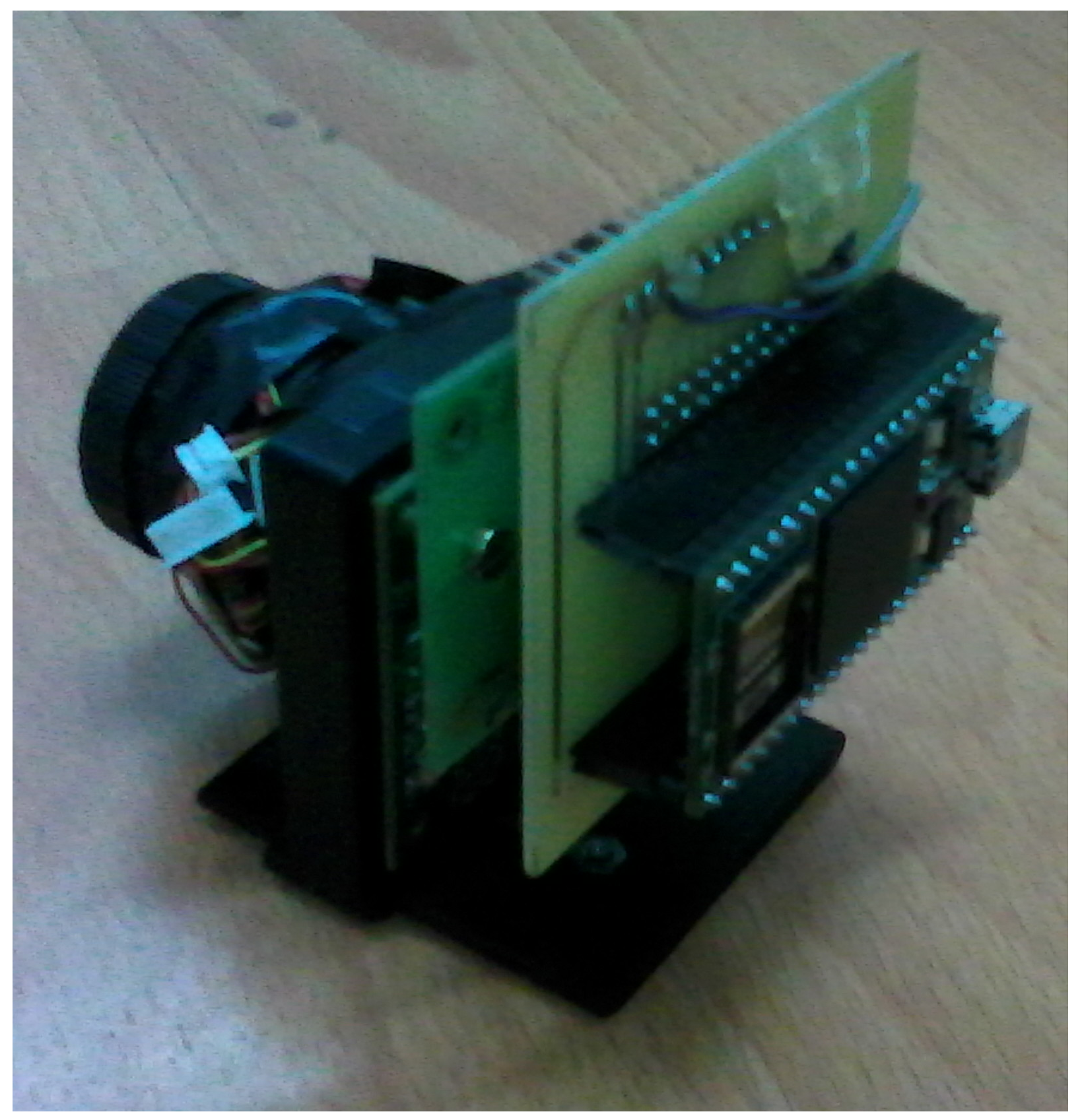}
%\hskip .1in
\includegraphics[height=60mm]{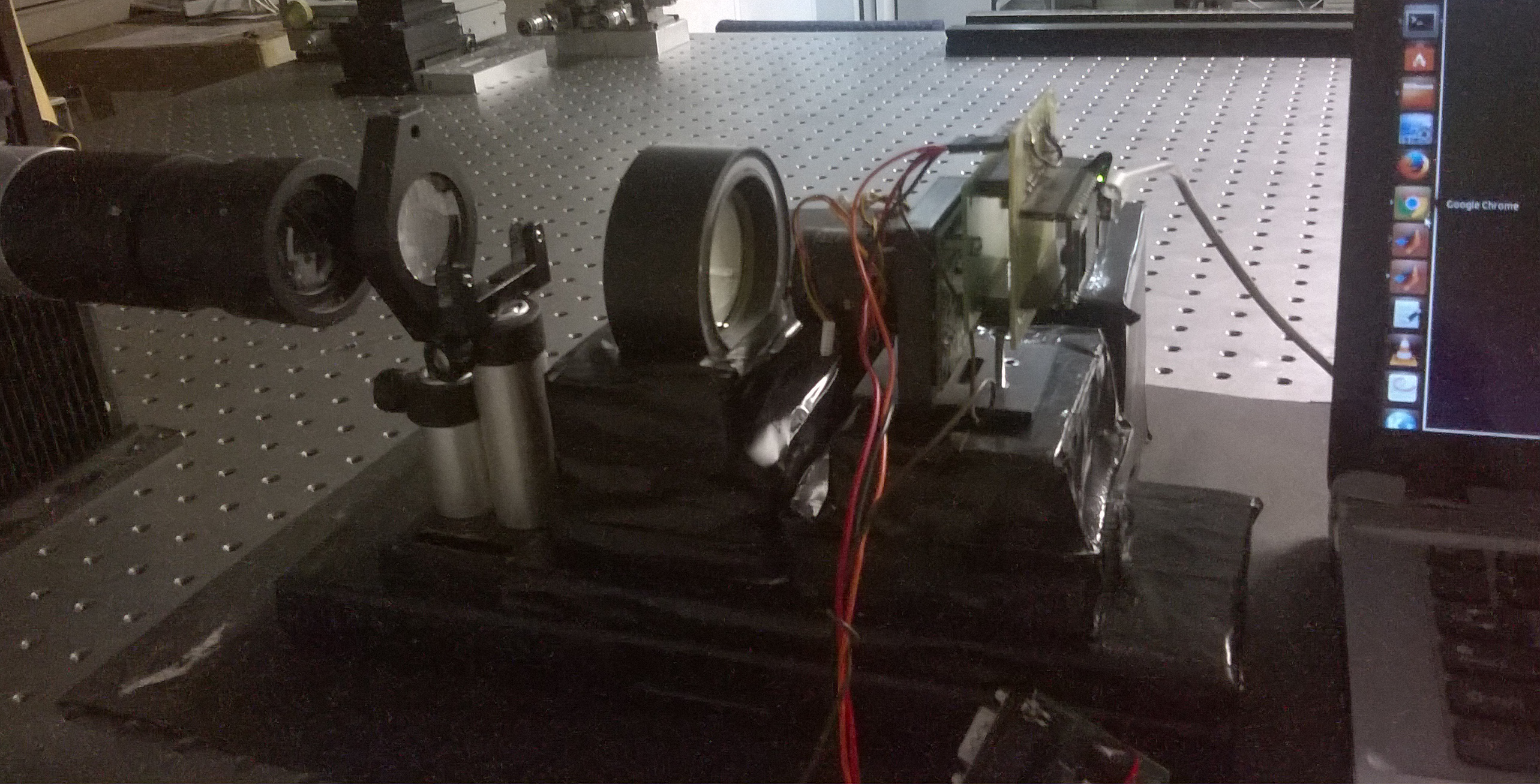}
% \end{minipage}
\caption{{\it Left}: Readout assembly of the detector. {\it Right}: Detector test setup on the optical table. From left to right: UV light source, collimating lens, and the  detector assembly with the MCP and the readout board.} 
\label{fig:photos}
\end{figure}

To perform the characterization of the whole detector, we assembled detector with the MCP, power supply board, relay optics, and the CMOS readout board on the optical table in a dark room (Fig.~\ref{fig:photos}, {\it Right}), with ambient temperature of $23^{\circ}$C. A faint uncollimated light source was provided in the form of a green LED with variable intensity, kept at a distance of about 10 cm from the MCP. 

\subsection*{Measurement of dark current}
\begin{figure}[H]
\begin{center}
\includegraphics[scale=0.4]{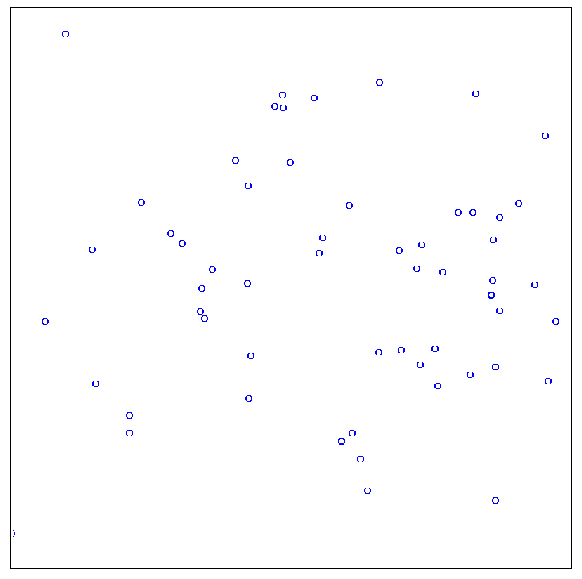} 
\end{center}
\caption{Dark count frame at $23^{\circ}$C (34 ms exposure). Highlighted are the detected dark counts.}
\label{fig:backgnd}
\end{figure}

We have done the basic tests in both photon counting and frame transfer mode, with the gain in the frame transfer mode lower than in photon-counting mode. The dark count test was performed by covering the detector window and reading the image of the MCP surface on the CMOS in photon counting mode. At room temperature, we saw a dark count of 1000 cps which is expected of an S20 photocathode. The dark count frame is shown in Fig.~\ref{fig:backgnd}, where the events above the threshold are highlighted.

\subsection*{Estimation of event gain}

To estimate the throughput in photon counting mode, we used a faint light from the LED lamp. Though the lamp is not flux calibrated, its intensity can be varied. The CMOS operates at the maximum frame rate of 30 fps, while storage and transmission units run at 40 MHz. This means ideally we can have up to 22000 cps global event rate, without any significant dead time due to the electronics. But for our current lab tests, we are using an RS232 connection for transferring the data to the PC, which supports count rates of only up to 6000 cps. Under gain saturation, we have a total number of $\sim 4 \times 10^6$ photons reaching the CMOS sensor for each event, which translates to 900 ADUs for the CMOS sensor under the Gaussian profile of the event (these values come from the instruments data sheets). Due to the demagnification from the lens, a single pore is confined to 1/4th of a pixel, but each event can be spread across multiple pores depending upon the front end optics.

\begin{figure}[!ht]
\begin{center}
\includegraphics[width=100mm]{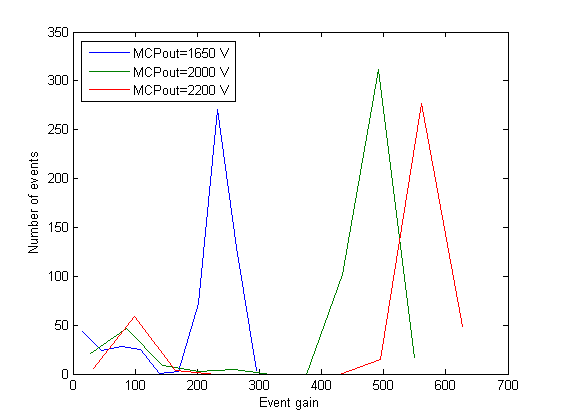}
\end{center}
\caption{{\it Left}: Histogram of event gains.}
\label{fig:gain}
\end{figure}

To better understand the gain characteristics of the detector, we have done the tests for gain variations by measuring the total CMOS ADUs under each event for a number of events due to a faint LED source. We measured the event gain by taking a histogram of the events for three different voltage settings. We can see from Fig.~\ref{fig:gain} that the actual event gain is less than the manufacturer's value, and that it varies considerably. For the cathode voltage of 2.2 kV (gain of $10^6$), we see a typical event gain of 550 ADUs instead of 900 ADUs, which may be due to the variations in the efficiency of the phosphor screen or the relay optics or both. 

\subsection*{Rejection of multiple events}

\begin{figure}[ht!]
\begin{center}
\includegraphics[scale=0.5]{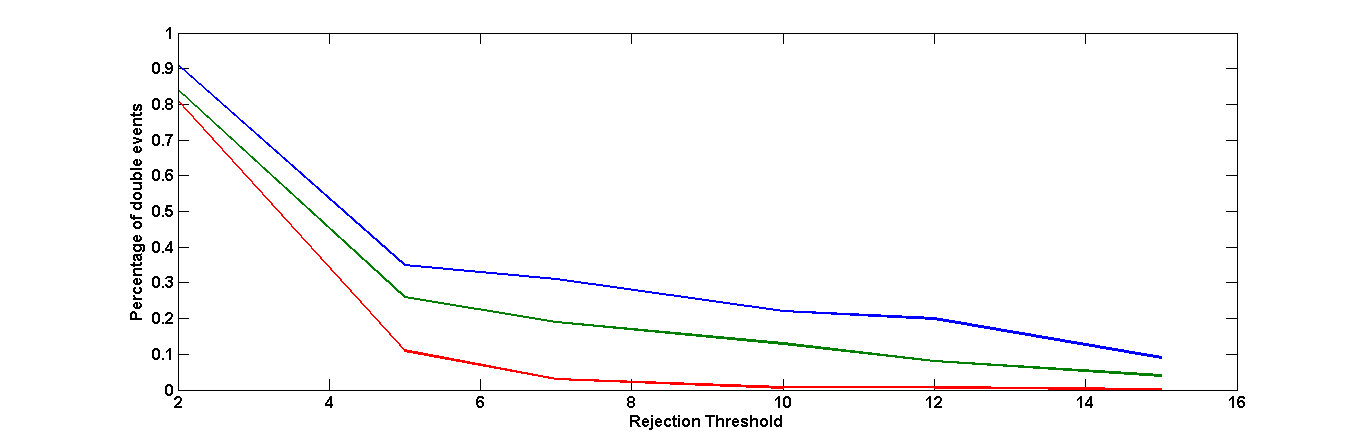} 
\end{center}
\caption{Double events vs rejection thresholds (in ADUs) for 3 different intensity levels of the LED lamp. Blue indicates the highest and red the lowest intensity.}
\label{fig:plot}
\end{figure}

Even with a very faint LED source, we see some double events in the detector. Such events are detected by an additional threshold value, which is the difference of maximum and minimum corner pixel values. For a $5\times 5$-window centroiding algorithm, we found out the number of double events that are remaining after applying a particular rejection threshold (Fig.~\ref{fig:plot}). In this figure, by different colors we mark the different levels of intensity of the LED lamp, with blue for the highest and red the lowest setting. Though we do not have calibrated flux, we can that at a higher rate of incoming photons we do see a large number of events marked as multiple events, which are detected by using an appropriate value of for the corner threshold. With fainter incoming flux, this number reduces fast, as expected. 

\subsection*{Imaging test}

\begin{figure}[!ht]
\begin{center}
\includegraphics[width=70mm]{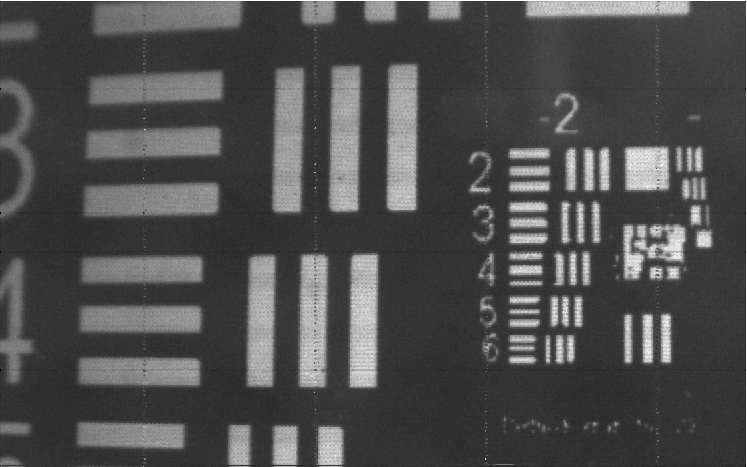} %100 percent
\hskip 0.3in
\includegraphics[width=70mm]{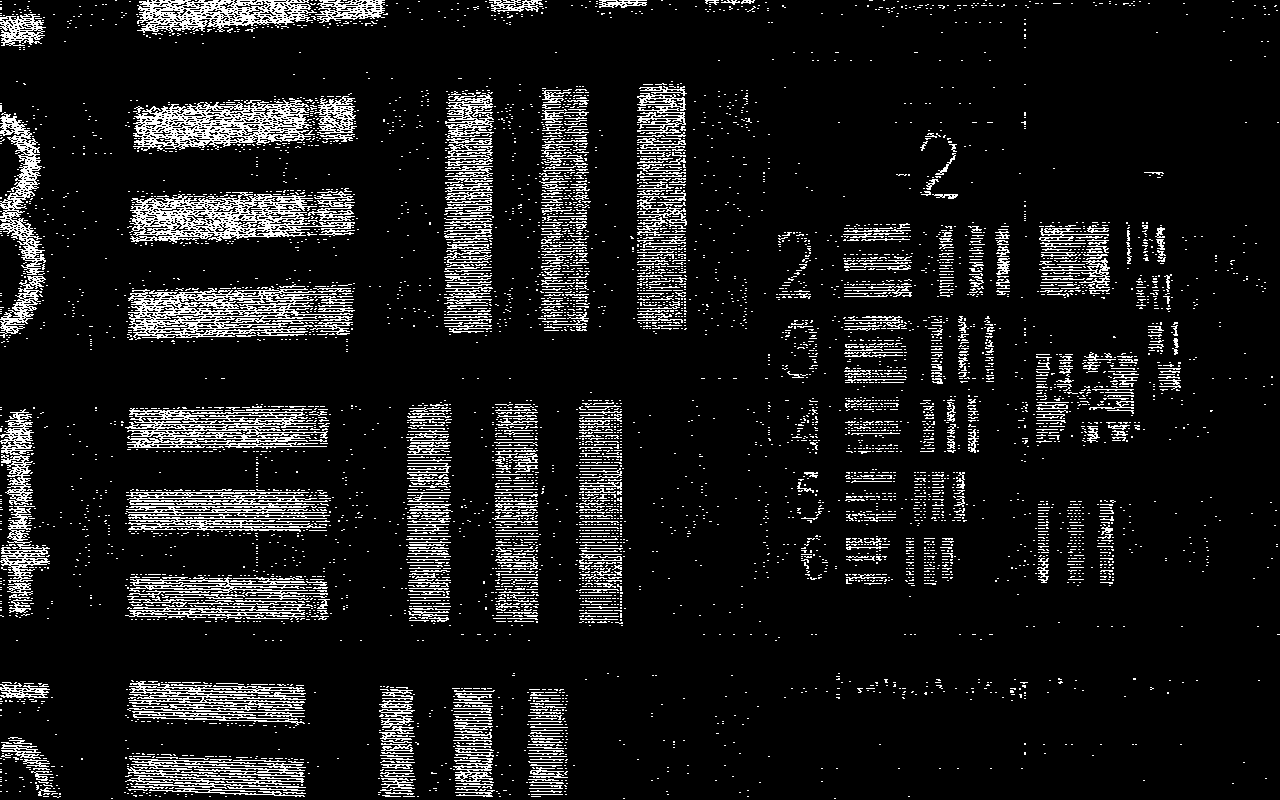}
\end{center}
\label{fig:image1}
\begin{center}
\includegraphics[width=70mm]{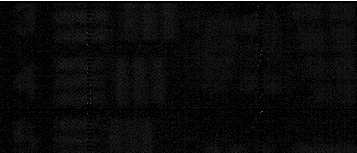} %100 percent
\caption{{\it Top Left}: Image of the USAF target in frame transfer mode. {\it Top Right}: Image reconstructed from the centroids. {\it Bottom}: Zoomed in version of the image reconstructed from the centroids (including the fractional parts)} 
\end{center}
\end{figure}

To test the overall performance of the detector, we took the images of a USAF target in both centroiding and frame transfer mode. For frame transfer mode, we set lower voltage values for the HVPS, and read the image off the MCP at the CMOS surface. The data rate was kept at 30 MBPS for a 24 MHz input clock rate in full frame, although we have transmitted only one frame per second to the host PC to account for the large transmission delays. The image shown in Fig.~\ref{fig:image1} ({\it Left}) is reconstructed by the pixel values read out serially from the sensor by a  MATLAB\textsuperscript{\textregistered} program.  The image in Fig.~\ref{fig:image1} ({\it Right}) is reconstructed from the centroids with a dynamically estimated threshold value. The data packets, shown in Fig.~\ref{fig:row3}, were read through the serial port by another MATLAB\textsuperscript{\textregistered} program. The centroids for the same frame ID were added together to form the image. The centroid values are accurate up to 1/16th of a pixel, though we have used only the integer parts for the reconstruction. In the current test set-up the resulting spatial resolution was rather low, 1.78 lp/mm, therefore we are not able to resolve within the single MCP pore. One of the reasons we are getting low spatial resolution is the performance of the relay lens (Sec 2.2). We are in the process of optimizing this. In addition, the incoming light was not perfectly collimated which could have introduced the loss of resolution.

We are still in the process of optimizing the interface between the MCP and the CMOS in order to improve the image quality, by using the new custom-made optics. To improve the current temporal resolution (30 cps) to handle faster event rates, we would need to re-program the internal registers of the CMOS through I2C interface, which is time consuming and difficult to implement in real time on the flight. Therefore, in the final version of the detector, we will replace the existing CMOS sensor with a high speed CMOS sensor NOIL2SM1300A from OnSemiconductors, that supports up to 500 fps at full frame and enables on-the-fly addressing of sub-regions.  

\section{Conclusions and Future Work}

In the present work, we have described an intensified CMOS detector with an FPGA-based readout developed in-house which can work as a camera in bright light, transferring each frame serially to the data processing system, and as a photon-counting detector with on-board centroiding in faint light conditions. The performance of the electronics is predominantly limited by the performance of the FPGA chip and the speed of CMOS data transfer.

We are now readying the detector assembly for use in an astronomical payload on small satellites, such as e.g. PISAT\footnote{PISAT is a 3-axis stabilized imaging satellite developed by PES University, Bangalore, India.  http://pes.edu/pisat/} {\cite{pisat}}. We also have a high-altitude balloon program for testing flight hardware \citep{Balloon_Sreejith}. The detector described in this work will be flown as the backend instrument in the near-UV spectrograph \cite{NUV_Spectrograph_Sreejith} on a high-altitude balloon experiment. 

\section{Acknowledgments}

We wish to thank Mr S. Sriram of the Indian Institute of Astrophysics for his valuable suggestions and for his help in setting up the tests. Part of this research has been supported by the Department of Science and Technology (Government of India) under Grant IR/S2/PU-006/2012.

\bibliography{bibfile}{}
\bibliographystyle{ws-jai}

% \newpage
% \appendix{Flow charts of algorithms}

% The flowchart (Fig.~\ref{fig:flowchart}) shows the program flow for the centroiding algorithm (Section 4.2). 
% \begin{figure}[ht!]
% \begin{center}
% \includegraphics[scale=0.6]{A1_Flowchart.jpg} 
% \end{center}
% \caption{Centroiding Flowchart}
% \label{fig:flowchart}
% \end{figure}

% The flowchart (Fig.~\ref{fig:flowchart}2) shows the program flow for the image reconstruction (Section 4.4).

% \begin{figure}[ht!]
% \begin{center}
% \includegraphics[scale=0.6]{A2_Flowchart2.jpg} 
% \end{center}
% \caption{Image Reconstruction Flowchart}
% \label{fig:flowchart2}
% \end{figure}

\end{document}